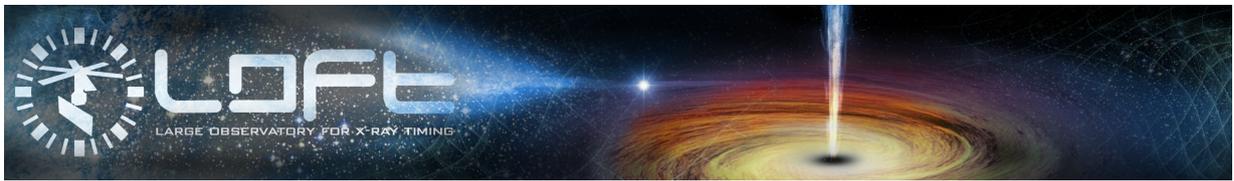

# Stellar flares observed by *LOFT*: implications for the physics of coronae and for the "space weather" environment of extrasolar planets

## White Paper in Support of the Mission Concept of the Large Observatory for X-ray Timing

### Authors

S.A. Drake[1], E. Behar[2], J.G. Doyle[3], M. Güdel[4], K. Hamaguchi[5],
A.F. Kowalski[6], T. Maccarone[7], R.A. Osten[8], U. Peretz[2], S.J. Wolk[9]

[1] USRA/CRESST and NASA/GSFC, Greenbelt, Maryland, USA
[2] Technion–Israel Institute of Technology, Haifa, Israel
[3] Armagh Observatory, Armagh, United Kingdom
[4] University of Vienna, Vienna, Austria
[5] UMBC/CRESST and NASA/GSFC, Greenbelt, Maryland, USA

[6] University of Maryland Department of Astronomy and NASA/GSFC, Greenbelt, Maryland, USA
[7] Texas Tech University, Lubbock, Texas, USA
[8] Space Telescope Science Institute, Baltimore, Maryland, USA
[9] Harvard-Smithsonian Center for Astrophysics, Cambridge, Massachusetts, USA

## Preamble

The Large Observatory for X-ray Timing, *LOFT*, is designed to perform fast X-ray timing and spectroscopy with uniquely large throughput (Feroci et al. 2014). *LOFT* focuses on two fundamental questions of ESA's Cosmic Vision Theme "Matter under extreme conditions": what is the equation of state of ultra-dense matter in neutron stars? Does matter orbiting close to the event horizon follow the predictions of general relativity? These goals are elaborated in the mission Yellow Book (`http://sci.esa.int/loft/53447-loft-yellow-book/`) describing the *LOFT* mission as proposed in M3, which closely resembles the *LOFT* mission now being proposed for M4.

The extensive assessment study of *LOFT* as ESA's M3 mission candidate demonstrates the high level of maturity and the technical feasibility of the mission, as well as the scientific importance of its unique core science goals. For this reason, the *LOFT* development has been continued, aiming at the new M4 launch opportunity, for which the M3 science goals have been confirmed. The unprecedentedly large effective area, large grasp, and spectroscopic capabilities of *LOFT*'s instruments make the mission capable of state-of-the-art science not only for its core science case, but also for many other open questions in astrophysics.

*LOFT*'s primary instrument is the Large Area Detector (LAD), a 8.5 m$^2$ instrument operating in the 2–30 keV energy range, which will revolutionise studies of Galactic and extragalactic X-ray sources down to their fundamental time scales. The mission also features a Wide Field Monitor (WFM), which in the 2–50 keV range simultaneously observes more than a third of the sky at any time, detecting objects down to mCrab fluxes and providing data with excellent timing and spectral resolution. Additionally, the mission is equipped with an on-board alert system for the detection and rapid broadcasting to the ground of celestial bright and fast outbursts of X-rays (particularly, Gamma-ray Bursts).

This paper is one of twelve White Papers that illustrate the unique potential of *LOFT* as an X-ray observatory in a variety of astrophysical fields in addition to the core science.

# 1 Introduction

*LOFT* can contribute greatly to our understanding of stellar flares. Expanding our knowledge of stellar flaring is crucial to examining the influence of transient sources of ionizing radiation from a host star on exoplanet systems, important for habitability concerns and space weather which other worlds might experience. Such studies extend the solar-stellar connection in determining the extent to which solar models for conversion of magnetic energy into plasma heating, particle acceleration and mass motions apply to the more energetic stellar flares. Extrapolating from lower efficiency high energy monitoring observatories, *LOFT* will detect at least 100 flares per year from stars exhibiting the extremes of magnetic activity. The anticipated source types include young stars, fully convective M dwarfs, and tidally locked active binary systems. The wide-field monitoring capability of *LOFT* will also likely enable the detection of flares from classes of stars not hitherto systematically studied for their flaring, and will be important for expanding our understanding of plasma physics processes in nondegenerate stellar environments. Three important questions that *LOFT* can address are:

- *What are the properties of the nonthermal particles responsible for the initial flare energy input?* The study of their distribution with energy and of their temporal behavior, e.g., spikes, oscillations, etc., will be enabled by *LOFT*'s sensitive hard X-ray capability, far in excess of previous missions.

- *What are the physical conditions of the thermal plasma whose emission dominates the later stages of stellar flares?* Through its soft X-ray capabilities, *LOFT* will enable us to study the variation of properties such as the temperature and elemental abundances as a function of time during flares.

- *What is the maximum energy that stellar flares can reach?* The wide-field, broad X-ray band and high-sensitivity capabilities of *LOFT* will enable us to detect the rare "superflares" that have been only rarely detected by previous and current missions: better understanding of the prevalence of such powerful events is important both for the physics of stellar flares and for the implications of their effect on potential planets orbiting these stars.

# 2 Solar flares

The first observation of a solar flare was made in visible ("white") light near a sunspot group in 1859 by two amateur astronomers (Hodgson and Carrington). The intense geomagnetic storm which occurred 18 hours later, as particles from the associated coronal mass ejection (CME) flooded into the earth's magnetosphere, was (correctly) hypothesized to be a by-product of this event. Almost a century later, Chubb et al. (1957) detected X-ray emission associated with a minor solar flare and suggested that it could affect the earth's ionosphere. Since then, for at least the last several decades, the X-rays from the Sun in general, and from solar flares in particular, have been extensively studied by the *GOES* satellites and other dedicated missions, and it has come to be realized that the emission in this energy band is strongly related to the fundamental underlying physical flare mechanism.

Due to our ability to spatially resolve structures the X-ray emission from the solar corona and flares, much has been learned about the flare mechanism. Flares occur in close proximity to active regions (ARs), which are effectively intense regions of localized kG-strength magnetic field. Loops from these ARs extend into the solar corona and, as the footpoints of these loops are jostled by solar convective motions, they are twisted and distorted until magnetic reconnection occurs. There is then a sudden release of energy, resulting in the acceleration of electrons and ions in these loops up to MeV energies, emitting nonthermal radio (gyrosynchrotron) and nonthermal hard X-ray emission (bremsstrahlung and Compton backscattering). These energetic particles stream both away from the Sun and also down the loop footpoints where they deposit substantial energy to the lower solar atmosphere (the chromosphere), producing the observed intense hard X-ray emission at the footpoints.

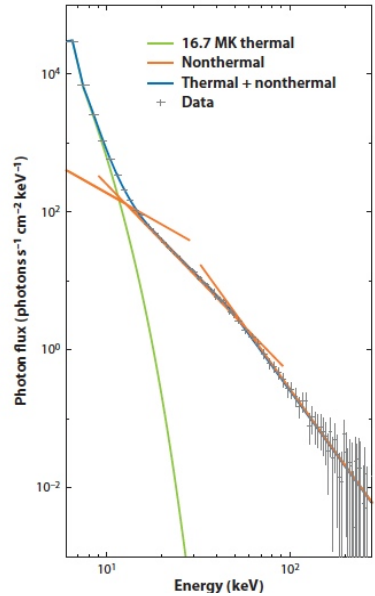

Figure 1: X-ray spectrum of the impulsive phase of a solar flare as observed by RHESSI, showing the thermal component which dominates the soft X-ray band below ~15 keV and the nonthermal component at higher energies (taken from Benz & Güdel 2010).

Thus, the initial 'impulsive' phase of a flare is an inherently nonthermal process: in particular, observations in the 20–1000 keV range of this phase show a power-law energy spectrum (actually, in general a broken power-law spectrum) which can be related to the power-law spectrum of the nonthermal electrons in the context of the standard "thick target" bremsstrahlung flare model (see Fig. 1). As the high-energy particles deposit their energy in the chromosphere, they heat and therefore "evaporate" plasma from this region to fill the flaring loop(s) with soft X-ray emitting plasma to temperatures of initially ~1–3 keV (12–40 MK) (in the solar case).

The hard X-ray emission observed in solar flares can "exhibit significant fluctuations on time scales as short as tens of milliseconds" (Kiplinger et al. 1983). The soft X-ray emitting flaring loop(s) cools on timescales of minutes to hours (for the largest flares) back to the typical AR coronal temperature of 0.2–0.3 keV (2–4 MK). It is this "decay" phase of the flare that produces the thermal emission which dominates the soft X-ray energy band. Quite frequently smaller flares occur during the decay phase of a large solar flare, usually from the same AR, but occasionally from another AR on the solar surface. Spatially resolved observations of solar flares in both the hard and soft X-ray bands show quasi-periodic pulsations (QPPs) which Foullon et al. (2005) interpret as being due to MHD kink-mode oscillations in the flaring loop(s). An additional source of X-ray emission during the impulsive and peak phase of the flare is caused by irradiation of the stellar surface by the thermal X-ray source and the resultant emission of fluorescent emission when these photons encounter the chromosphere, with a significant Fe K-alpha 6.4 keV emission line being produced. The largest solar flares observed in the modern era have peak soft X-ray luminosities of $\sim 10^{28}$ erg s$^{-1}$ and inferred total radiative emitted energies of $\sim 10^{32}$–$10^{33}$ erg.

On the present-day Sun, a middle-aged star of relatively low activity, the rate at which flares occur significantly varies with the solar 11-year activity cycle. At solar maximum, there may be a few large ARs and many more small ARs on the solar surface at any one time, and the flare rate can be dozens per day. At times during solar minimum, the solar surface may be devoid of ARs, and no significant flares at all may be detected for days.

## 3 Stellar flares

Stellar flares have been studied for over 6 decades since the discovery of white light flares from M dwarfs in the 1940s. When gamma-ray bursts (GRBs) were discovered in the 1960s, one of the early hypotheses of their nature was that they were the hard X-ray counterparts of stellar flares, but the lack of optical counterparts or spatial coincidence with nearby active stars soon made this idea untenable (although in fact some stellar flares have triggered GRB detectors, e.g., the 2008 April 25 flare of the dMe star EV Lac observed by *Swift* was also detected by the Konus/Wind GRB detector in the 18–70 keV band).

The X-ray emission from stellar flares was first detected by the *ANS* satellite in 1975, which detected soft (∼0.3 keV) X-ray emission from the dMe star UV Ceti during a large optical flare. The active 2.9 day period eclipsing binary system Algol (B8V + K2IV), was later detected as an X-ray source in the 2–6 keV band by the *SAS-3* satellite in 1975, likely during a small flare, and, since then, essentially every major X-ray mission from *Ariel-V*, *HEAO-1*, *Einstein*, etc., has detected stellar flares. It was soon realized that the X-ray luminosities of many of these stellar flares were orders of magnitude larger than those of the largest ever observed solar flares ($\lesssim 10^{28}$ erg s in the 0.1–100 keV band). The observed X-ray fluxes of stellar flares have been as high as $10^{-8}$ erg cm$^{-2}$ s$^{-1}$, with corresponding peak X-ray luminosities as high as $10^{32}$ erg s$^{-1}$ for nearby dMe stars (which can exceed their normal bolometric luminosities!) , and $10^{33.5}$ erg s$^{-1}$ for active (RS CVn and Algol-type) binaries in the solar neighborhood. Flares from young ($< 10$ Myr) stars in star-formation regions such as Orion or in nearby young stellar associations such as the TW Hya group have been observed to reach X-ray luminosities as high as $10^{33}$ erg s$^{-1}$ (Getman et al. 2008; Morii et al. 2010), albeit with somewhat lower fluxes at the Earth due to their greater distances.

In contrast to the Sun with its activity cycle, active stars, such as the dMe stars and active binaries, appear to be essentially always "on" (although Richards et al. 2003 found evidence that radio flaring in some RS CVn and Algols is modulated on a super-period 10–20 times longer than their rotational periods) and their (apparently) "quiescent" coronae are likely the result of continuous low-level flaring. The evidence for this comes from (1) optical, (2) X-ray, and (3) radio observations: (1) their optical spectra and photometry show evidence for the persistent presence of ARs (starspots and Ca II emitting plages), (2) even without obviously flaring their coronae have a significant emission measure at $T \sim 10$–30 MK, and, in some cases, the rate of X-ray flares show a power-law distribution with a slope of $\sim -2$ to $-2.5$ indicating that "the bulk of the energy release by flares comes from the large number of small events that may not be resolved individually" (Benz & Güdel 2010); this fundamental role of flares in the heating of a million-degree corona (Doyle & Butler 1985) has in fact equally been proposed for the Sun itself albeit at levels scaled-down by several orders of magnitude from to the active-stellar case; the hypothesis has come to be known as the "microflare" or "nanoflare" hypothesis (e.g., Parker 1988), (3) the nearest active stars are persistent bright nonthermal gyrosynchroton emitters at cm wavelengths, indicating that, unlike the Sun, particles are being accelerated to MeV energies effectively all the time. One general property of both solar and stellar flares is that the larger the flare, the hotter the peak temperature reached in the flare. The largest stellar flares observed from active stars have had peak temperatures in excess of 100 MK, e.g., the 1997 flare from Algol observed by *BeppoSAX* (Favata & Schmitt 1999). For young stars which are still actively accreting another type of flare has been proposed, where the reconnection may occur in magnetic loops which extend from the stellar surface to the surrounding accretion disk rather than back to the stellar surface.

Given what we know about the hard X-ray emission of solar flares, particularly during the impulsive rise and peak phases, there have been a number of studies of stellar X-ray emission, particularly during flares, which have searched for the presence of a non-thermal power-law component. Given that in the solar case, it typically only starts to dominate the spectrum at energies $E > 10$–30 keV, it is not surprising that even the most sensitive soft X-ray telescopes in the 0.5–10 keV band, such as *Chandra* and *XMM-Newton*, have failed to find this power-law component. Previous broad-band observatories such as *Ginga*, *RXTE*, and *BeppoSAX* have had sufficient sensitivity to detect a small number of large flares from active stars at energies of ∼50 keV as previously mentioned, but in every case detected only very hot thermal emission.

The current *Swift* mission has observed ∼6 large stellar flares which have been bright enough ($\gtrsim 3 \times 10^{-9}$ erg cm$^{-2}$ s$^{-1}$ in the 15–50 keV band) to trigger the BAT and cause the satellite to slew to point its soft X-ray and UV/optical telescopes for detailed follow-up. In most cases only high-temperature ($T > 100$ MK) thermal emission was detected, e.g., the dMe stars EV Lac and DG CVn (Osten et al. 2010; Drake et al. 2014; Kowalski et al. 2015), but in one case (Osten et al. 2007), a flare from the RS CVn binary II Peg on 2005 Dec 16, an excess (over thermal) emission component was observed in the 40–∼100 keV range for a few hours after the flare onset. The authors interpreted this component as most likely the long-sought-after nonthermal emission component: a second possibility, that of an ultrahot (26 keV = 300 MK) component, was deemed unlikely on the grounds that such hot plasma could not remain confined by the stellar magnetic field for the observed duration (as discussed by Osten et al. (2010), the detection of the Fe K$\alpha$ 6.4 keV feature persisting into the decay phase of a flare is likely due to collisional ionization from the nonthermal particles rather than photoionization, and thus can also constrain the nonthermal energy spectrum). The typical durations of these *Swift*-detected flares is several minutes to several hours in the hard X-ray ($E > 15$ keV) band but much longer (12 hours to many days) in the soft X-ray band. The soft X-ray emission from the flare of DG CVn (classifiable as a superflare based on its $10^{36}$ erg total radiated energy) took about 2 weeks to fade to its normal non-flaring level (see Fig. 2): in fact, for several minutes at the flare peak, the soft X-ray flare luminosity exceeded DG CVn's normal bolometric luminosity. The left panel of Fig. 3 shows the simulated WFM 2–50 keV spectrum of the first 300 s of the DG CVn flare. If such a flare triggered a pointing in this direction by *LOFT*, Fig. 3 (right) shows what the LAD might detect $10^4$ s later (again based on the DG CVn flare).

Figure 2: The Swift XRT soft X-ray (0.3–10 keV) light curve of the flare of DG CVn on 2014 April 23 (blue points correspond to XRT data in WT mode, red points to data in PC mode; from Kowalski et al. 2015). The peak X-ray flux corresponds to $5 \times 10^{-9}$ erg cm$^{-2}$ s$^{-1}$.

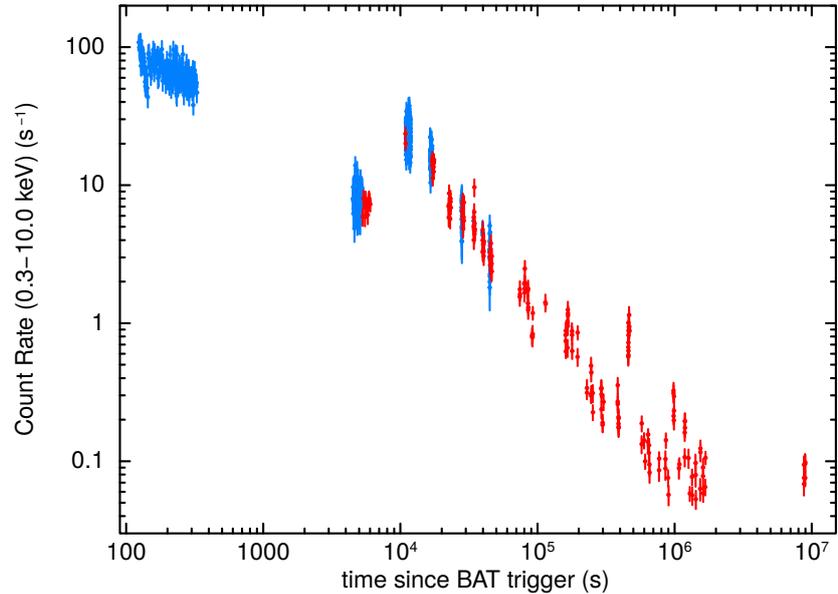

Doyle et al (1992) detected a large flare on the RS CVn star II Peg with a similarly high estimated radiated energy of $10^{36}$ erg, or approximately 0.1% of its available magnetic energy. For even larger energetic stellar flares, storage of energy in an intra-binary region may have to be considered (van den Oord 1988). Evidence for enhanced intra-binary emission has in fact been seen previously, e.g., Gunn et al. (1999).

The possibility of extremely large flares from "normal", i.e., inactive, stars similar to the Sun was first suggested by Schaefer (1989) and further pursued by Schaefer et al. (2000) in the context of older solar-mass stars. These authors presented a rather heterogeneous sample of proposed large flares taken from the literature, mostly based on optical observations. They defined "superflares" from such stars as flares having radiated energies $> 10^{33} - \sim 10^{34}$ erg, i.e., in excess of the energies of the largest solar flares observed in the modern era, and suggested an occurrence rate of 1/600 years. This hypothesis has received some support from a number of studies which have utilized the exquisite Kepler wide-band optical photometry for a large number of late-type stars. Shibayama et al. (2014) detected 1547 superflares from 279 G-type dwarfs in 500 days of Kepler data, and derived an average superflare occurrence rate in stars with rotational periods > 10 days of 1/800–5000 years.

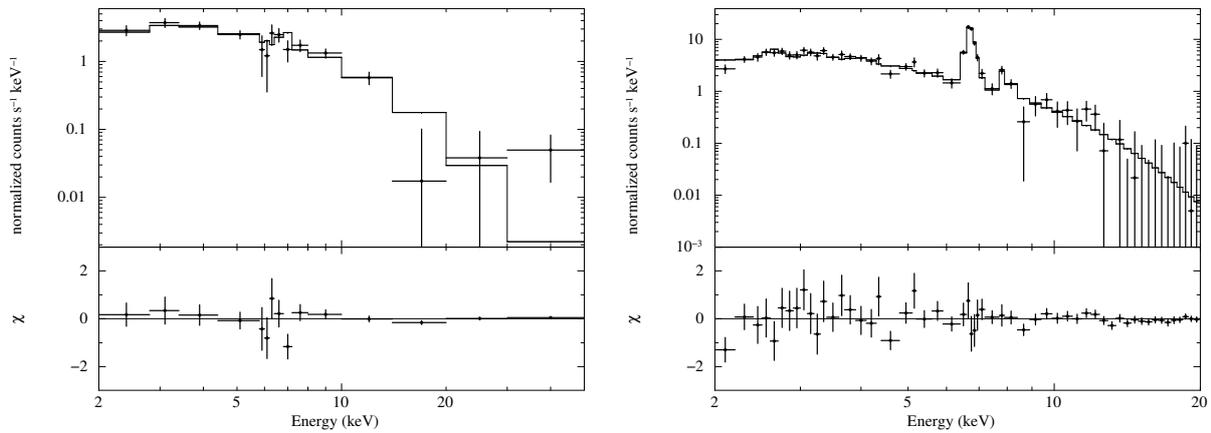

**Figure 3:** *Left:* Simulated *LOFT* WFM 16-channel spectrum of the first 5 minutes of a flare with similar flux and spectrum to the 2014 April 23 "superflare" of DG CVn. *Right:* Simulated *LOFT* LAD spectrum of a 1000 s time segment of a flare with similar flux and spectrum to the second DG CVn flare (the one at $t + 10^4$ s in Fig. 2) on 2014 April 23. Notice the well-exposed He-like Fe lines at 6.7 keV and 7.9 keV which would enable an accurate coronal Fe abundance to be derived.

Follow-up spectroscopic studies of a few of these stars (Nogami et al. 2014) appear to indicate that at least some these stars are similar to the Sun in age, rotation period and magnetic field properties.

## 4 Impact of stellar flares on planetary companions

The frequency of occurrence and maximum energies of stellar flares also has implications for the probable effects of their UV and X-ray emission, high-energy particles and associated CME on any planets that might be orbiting the stars. As noted above, the Carrington flare in 1859 produced a major geomagnetic storm and damaged the nascent telegraphy network: it has been estimated that a similar storm nowadays could cause one trillion of euros damage to communication systems, power grids, satellites, etc. There were storms of somewhat smaller magnitude in 1972 and 1989 (this latter caused $0.5 billion damage to the Quebec power grid), and there have been other similarly large solar flares whose location on the Sun has been such that their associated CMEs missed the Earth.

The effects on the Earth's atmosphere and biosphere (or of an earth-sized planet in the habitable zone of its star) of a superflare orders of magnitude greater than the Carrington event would clearly be much more substantial, and, in the worst case scenario, result in a mass extinction. Planets in the habitable zone of M dwarf stars ($\sim 0.1$ AU from the star) would fare even worse due to their closer proximity, e.g., the 2014 flare of DG CVn produced a radiative flux on such a hypothetical planet $10^6$ times greater than that which the Carrington flare produced on the Earth! M dwarfs also take much longer to spin down than solar-mass stars and thus can maintain their high levels of coronal activity and flaring for a very long time ($\gg 10^8 \ldots 10^9$ years): because of this, planets orbiting M dwarf stars are unlikely to develop life as quickly as it appeared on the Earth ($< 10^9$ years).

The role of X-ray flares in planetary atmospheric erosion of close-in planets was recently demonstrated for the hot Jupiter HD 189733b. In two HST/STIS observations of the Ly$\alpha$ profile, one showed excess absorption during transit, amounting up to about 12% at velocities of $-140$ to $-230$ km s$^{-1}$, while the other did not (Lecavelier des Etangs et al. 2012). The measured velocity exceeds the planet's escape velocity of 60 km s$^{-1}$. The increased level of hydrogen absorption in one of the observations coincides with the occurrence of a large X-ray flare recorded by *Swift* a few hours before transit. Radiation pressure from the star is insufficient to accelerate hydrogen atoms to the observed speed, although charge exchange with stellar wind protons may work, in conjunction with an EUV flux five times the present solar value controlling the hydrogen ionization timescale. Alternatively, the high level of X-rays may lead to the observed level of evaporation (i.e., an extended neutral hydrogen cloud).

## 5  What is the key science that *LOFT* will address?

1. **A much increased rate of detection of large flares from nearby active stars.** *LOFT* 's capability of detecting soft X-ray transients over a wide area of the sky will be much superior to previous and current missions with wide-field X-ray monitors, such as MAXI and *BeppoSAX*. Based on the results of 5.2 years of MAXI/GSC monitoring of the sky in the 2–10 keV band, we can estimate the expected frequency of flares from nearby ($\lesssim 50$ pc) active stars which *LOFT* will detect. From a search of ATELs issued by the GSC team, there are a total of 21 such flares above $10^{-9}$ erg cm$^{-2}$ s$^{-1}$ which have been reported, i.e., a rate of four per year. About 65% of the stars which flared were nearby evolved active RS CVn and Algol binaries (e.g., Algol, HR 1099, UX Ari), with the remaining 35% being nearby active M dwarf stars (e.g., YZ CMI, CC Eri, V1054 Oph) and one nearby T Tau star (TWA 7). MAXI scans the sky such that a given position is in the GSC FOV for ~45 s once every 92 m orbit, i.e., its duty cycle is ~1%. Most of the M dwarf flares were detected only once, implying durations < 1.5 h, whereas the active binary flares typically lasted hours to days. In contrast to MAXI, the *LOFT* WFM will operate by monitoring one third of the sky all the time, so that, based on the MAXI detections, we expect that it will detect > 100 flares yr$^{-1}$ above the same threshold flux, which is achievable at a $5\sigma$ level in a standard WFM 5 minute integration. A similar estimate can be obtained based on the number of stellar flares detected by the *BeppoSAX* Wide Field Camera as reported by Heise et al. (2000). In Fig. 5, we show the flare detection rate as a function of flare energy for M and G dwarfs (Ramsay and Doyle 2015).

2. **The potential detection of flares from classes of stars not previously confirmed (at least in the X-ray band) of being capable of producing large flares.** For example, if solar-type dwarf stars do exhibit the claimed superflare rate of one per ~1000 yrs as discussed above, given that there are ~3700 G V stars within 50 pc of the Sun, *LOFT* might detect ~4 superflares per year from this group (see Fig. 5). Several types of giant stars have been shown to produce extremely energetic flares but these remain poorly studied because of their very long timescales. Giants such as $\beta$ Cet (Ayres et al. 2001) or $\sigma$ Gem (Osten 2002) have shown flares with time scales of several days in long EUVE observations but such events would be difficult to recognize in short (< 1 day) observations typical of modern X-ray observatories. The class of rapidly rotating FK Com type giants have some of the most energetic and hottest coronae known observed so far, and have also been demonstrated to produce among the most energetic flares ever observed at radio and optical wavelengths (e.g., Bunton et al. 1989, who derived an integrated microwave luminosity for HD 32918 corresponding to 1000 times the average solar X-ray luminosity), but their flare timescales of a few weeks have made extended X-ray monitoring difficult to impossible to accomplish. The *LOFT* WFM will be ideal for this kind of long-term study.

3. **Detection of chemical composition changes during large flares**. Chemical changes have been detected during large flares of X-ray active stellar coronae using both *Chandra* and *XMM-Newton* grating spectrometers. Elemental abundances during flares vary based on the First Ionization Potential (FIP) when compared to quiescence (e.g., Nordon & Behar 2007, 2008).The high photon count rate on the *LOFT* LAD detector will make up for a lower spectral resolution, in particular at high energies in the Si, S, Ar, and Fe K bands. This allows for accurate abundance measurements for hot flares (> 1 keV), with FIP trends clearly discernible (Fig. 4). We simulated typical abundance variations in a 1 keV and 2 keV flare, for a moderate flux of $5 \times 10^{-12}$ erg cm$^{-2}$ s$^{-1}$ and a 5 ks exposure. These fluxes are conservative for bright coronal X-ray sources, especially during flares. Initially solar values were assumed, and during the flare high FIP abundances (S, Ar) were doubled (the flux remains $5 \times 10^{-12}$ erg cm$^{-2}$ s$^{-1}$) while low FIP elements (Si, CA, Fe) were held at solar values. It is clear that the abundance changes are readily detected with the *LOFT* LAD. Fits on simulations of 2-keV plasmas constrain abundances well, with relative errors of S: ~15%, Ar: ~20%, Fe: ~10%, Ca and Si: ~40%. Higher fluxes than $5 \times 10^{-12}$ erg cm$^{-2}$ s$^{-1}$ would allow for even more precise measurements, with error on the abundances as low as a few percent for fluxes of order of $5 \times 10^{-10}$ erg cm$^{-2}$ s$^{-1}$.

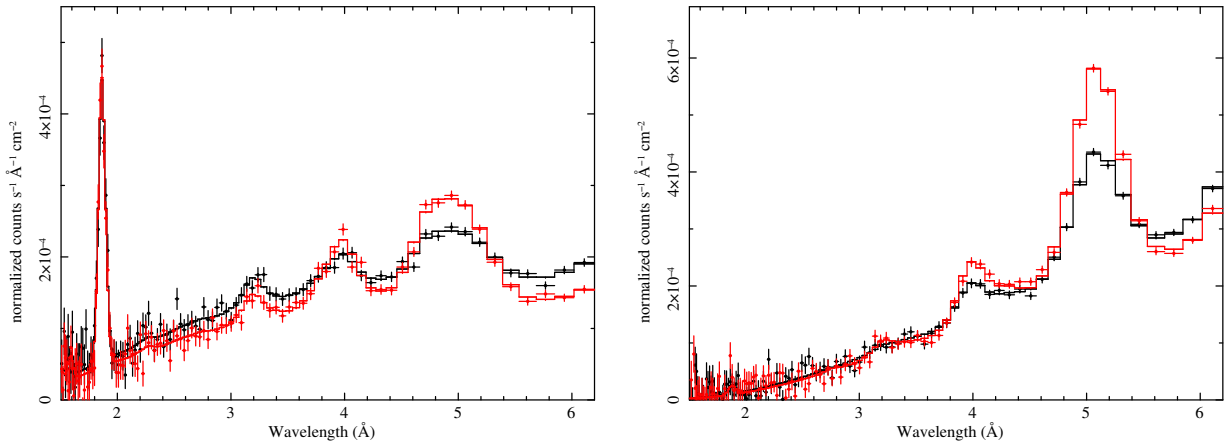

Figure 4: 5 ks simulations of a *LOFT* LAD observation of a high FIP increase (red data and model) for plasmas with temperatures of 2 keV (left) and 1 keV (right). Both simulations assume a flux of $5 \times 10^{-12}$ erg cm$^{-2}$ s$^{-1}$. K-shell lines of Fe (1.9Å), Ca (3.2Å), Ar (4.0Å), S (5.0Å), and Si (6.1Å) can be seen.

4. **The detection of the non-thermal emission and fluorescent Fe Kα from the early flare stages**. This can enable the derivation of the power-law index of the electrons producing the observed X-ray emission, in the context of the thick-target bremsstrahlung model (or other proposed models) for flares. Because of limitations of sensitivity, energy bandpass and response times of current and past X-ray observatories, there have been few such studies possible to date, with the Swift observation of the large flare of II Peg in December 2005 (Osten et al. 2007) being arguably the most convincing.

5. **The detection of short-timescale structure in the X-ray emission from stellar flares similar to the hard X-ray "spikes" and the X-ray QPPs which are seen in some solar flares**. Present instrumentation on current X-ray missions (mostly pointed telescopes rather than all-sky or wide-field monitors, except for Swift) has only started to address this issue, e.g., the X-ray oscillations reported in the dMe star AT Mic reported by Mitra-Kraev et al. (2005) or the 30–40 s oscillations as seen at ultraviolet wavelengths by Welsh et al. (2006) for several dMe stars. The detection of such fast-varying temporal structures at X-ray wavelengths by current and past missions requires the fortuitous detection of a large stellar flare in order to be statistically significant. In contrast, *LOFT*, with its huge effective area, will be able to search for these temporal signatures in the more frequent smaller flares that are typically seen to occur during long ($> 10^4$ s) "sit-and-stare" observations of active stars.

6. **Signatures of hard X-rays from the superposition of frequent flares**. Detection of continuous non-thermal hard X-rays could be a signature of continuous flaring in stellar coronae, and thus provide the elusive evidence for processes that have been suspected from "quiescent" radio emission. Detection of such emission could i) provide evidence for the "frequent flaring process", ii) explain the quasi-continuous acceleration of high-energy electrons seen in microwaves, and iii) indicate what energy distributions flares follow, including individually detected big flares (see Sect. 6 for more details).

7. **Detection of pre-main sequence (T Tau) star and protostar flares.** Most high- and low-mass star-formation regions are likely out of *LOFT*'s reach, given its sensitivity and the LAD's lack of any spatial resolution, but the brightest flares from stars in the nearer ones, such as the Tau-Aur, Oph and Cha star formation regions, and flares from even nearer isolated young stars, such as the TWA and $\beta$ Pic

Associations, will likely be detected. There indeed have been some reports of gigantic flares from T Tau stars (TTSs) of ~1 Myr age, in addition to the one detected by MAXI from TWA-7 (discussed above): for example, the TTS V773 Tau had a very luminous flare with a peak X-ray luminosity of $10^{33}$ erg s$^{-1}$, i.e., a flux of $2 \times 10^{-10}$ erg cm$^{-2}$ s$^{-1}$ at its distance of ~150 pc (Tsuboi et al. 1998). Thus, *LOFT* should be able to observe moderately bright ($10^{32}$ erg s$^{-1}$) X-ray flares from TTSs in the nearby associations, and the brightest flares ($10^{33}$ erg s$^{-1}$) from TTSs in nearby star formation regions. The nearest protostars (stars with ages ~0.1 Myr) are found in the Tau, R CrA and $\rho$ Oph star formation regions at 120–150 pc, so their flare luminosities would need to be $\gtrsim 10^{33}$ erg s$^{-1}$ to be detectable by *LOFT*. There are only a few detections of such large flares from protostars, e.g., a flare from YLW 15 ($L_{X,\text{peak}} \sim 10^{34} \ldots 10^{36}$ erg s$^{-1}$) detected by the *ROSAT* PSPC (Grosso et al. 1997) and one from YLW 16A ($L_{X,\text{peak}} \sim 10^{31} \ldots 10^{32.7}$ erg s$^{-1}$) by the *ROSAT* HRI (Grosso 2001). However, these detections were in the very soft X-ray band, so the luminosity estimates required large and somewhat uncertain corrections for intervening absorption.

8. **Stellar flares simultaneously observed by *LOFT* and long-wavelength radio observatories and/or all-sky instantaneous optical observatories**. The availability in the time-frame of *LOFT* of wide-field low-frequency radio observatories such as the SKA (`https://www.skatelescope.org/radio-transients/`) and of wide-field optical facilities such as the LSST (`http://www.lsst.org/lsst/science/science_goals`) and Evryscope (`http://adsabs.harvard.edu/abs/2014SPIE.9145E..0ZL`) will revolutionize our understanding of the multi-wavelength properties of stellar flares and other types of X-ray transients. M dwarf star flares will likely be the dominant class given the ubiquity of low-mass stars. For example, the 2014 flare of the young M dwarf DG CVn observed by *Swift* produced an initial brightening in the V-band of 2.5 mag (recorded by the *Swift*/UVOT) and a brightening of > 50 in the 2 cm radio flux density to >100 mJy (recorded by the AMI telescope 600 s after the hard X-ray peak: see Fender et al. 2014). Multi-wavelength flare studies can be (and have been) used to derive information on flare physical parameters, the properties of the nonthermal electrons and the coronal magnetic field strengths, etc., which are unobtainable from X-ray data alone. For more information, see White et al. (2011), who review the extensive studies of solar flares done using simultaneous radio and hard X-ray bands, and Slee et al. (2014) for a recent example of such a multi-wavelength campaign which observed 3 flares from the nearby young star AB Dor.

## 6  Appendix: signatures of continuous flaring

In the picture of "microflare" heating, a large number of relatively small flares act together to heat the observed solar/stellar corona quasi-continuously. Only on the Sun can such processes be observed down to the crucial, extremely numerous but extremely weak nanoflares, thanks to spatially resolved observations. On active stars, indirect evidence discussed further above suggests an important role of flares in coronal heating. The detection of numerous active stars in the hard X-ray band by the *LOFT* WFM with simultaneous observations of the thermal spectrum in the soft band provides another, much more direct access to frequent flaring in parallel to the observation of continuous non-thermal radio emission from such stars (we note that the latter provides an important tracer of such processes but is energetically unimportant; the hard X-ray band gives much more insight into the bulk energy release). Our hope is to find a hard X-ray component in the spectrum that is incompatible with the extrapolation of the lower-energy thermal X-rays (which will also be observed by *LOFT*).

Solar and stellar X-ray and U-band observations have established that large samples of flares are distributed in total radiated energy (or also flux amplitude) following a power-law,

$$\frac{dN}{dE} = kE^{-\alpha} \tag{1}$$

where $dN$ is the number of flares per unit time with a total radiated energy in the interval $[E, E + dE]$, and the power-law index $\alpha$ determines the relative role of smaller and larger flares (for efficient microflare heating, $\alpha$ exceed a value of 2).

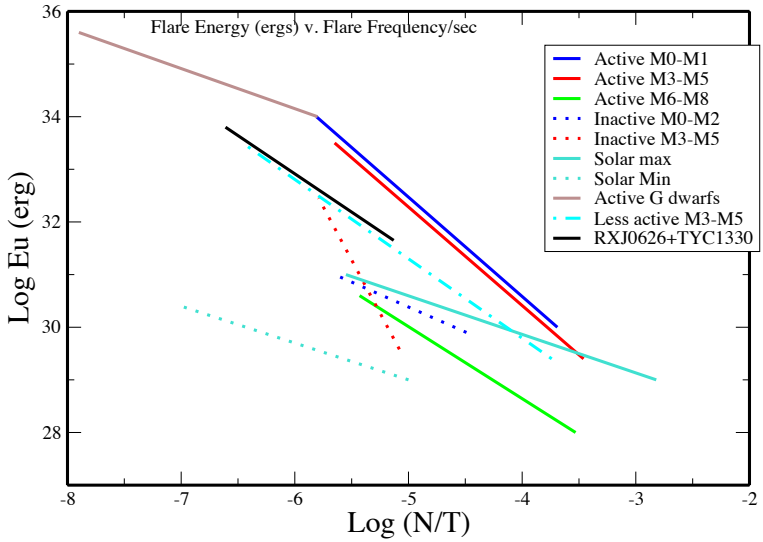

Figure 5: The cumulative flare frequency (in s$^{-1}$) versus U-band flare energy (in ergs) for G and M-type dwarfs. M dwarf data compiled from Moffett (1974), Doyle & Mathioudakis (1990), Dal & Evren (2011), Hilton (2011), and Hawley et al. (2014). Solar data compiled from Shibayama (2013) and Kretzschmar (2011). G dwarf data from Shibayama et al. (2013). See Ramsay & Doyle (2015) for more details.

Such distributions have been frequently reported for the Sun (e.g., Crosby et al. 1992) but are also known for several magnetically active stars (e.g., Audard et al. 2000, Güdel et al. 2003). Related distributions have been found in the optical range, as also reported here (Fig. 5 for cumulative distributions). Because stellar distributions refer to soft X-rays or extreme-ultraviolet radiation, conversion to the hard X-ray emission level from the flare distribution (Güdel et al. 2009). For typical observed parameters, we derive $L_{hard} = 8.5 \times 10^{-7} L_X$ erg keV$^{-1}$ at 35 keV, where $L_X$ is the average, continuously observed soft X-ray luminosity of the star. An approximate scaling to a wider band can easily be done; a factor of 50 higher broad-band flux above 20 keV should be expected, i.e., $L_{hard,tot} = 4 \times 10^{-5} L_X$. Should a systematic excess of hard X-rays of this kind be identified collectively in a large sample of active stars observed by *LOFT*, then this may be an important indicator – like nonthermal radio emission – for frequent flaring and particle acceleration. The spectral excess may, in conjunction with the power-law slope value, provide important consistency checks with other indicators for frequent flaring and thus further support (or weaken) the hypothesis of coronal heating by flare processes.